# Quantifying Robustness to Unmeasured Confounding in Time-Varying Treatment Confounder Settings: An Extension of E-value Approach

**Authors:** Md. Niamul Islam Sium[1*]

**Affiliations**

[1] Institute of Statistical Research and Training, University of Dhaka, Dhaka, Bangladesh

**\* Corresponding author**

Md. Niamul Islam Sium (email: nsium@isrt.ac.bd)

# Quantifying Robustness to Unmeasured Confounding in Time-Varying Treatment Confounder Settings: An Extension of E-value Approach

## Abstract

**Background:** The E-value has become widely used for assessing robustness to unmeasured confounding in observational studies, but the original framework was developed for single time-point exposure-outcome settings. This study extends the E-value methodology to longitudinal set up with time-varying treatments and confounders, where treatment-confounder feedback occurs.

**Methods:** A combined bias factor accounting for unmeasured confounding at multiple time points was extended, with three reporting scenarios presented: equal bias distribution across time points, confounding at a single time point, and a general case visualizing all possible confounder strength combinations.

**Results:** In simulations with an observed risk ratio of 1.73, unmeasured confounders with 1.96-fold associations at each time point could nullify the effect under equal distribution-substantially lower than the single time-point E-value of 2.85. Re-analysis of a published insulin resistance and cardiovascular disease study yielded similar patterns, with time-varying E-values of 1.63 at each time point compared to the originally reported 2.09.

**Conclusions:** Studies more like longitudinal set up may be more vulnerable to unmeasured confounding than single time-point E-values suggest. This extension provides accessible tools for transparent sensitivity analysis in time-varying settings while preserving the simplicity and minimal assumptions that make E-values widely applicable.

## Introduction

Estimating causal effect from observational data is a challenge in epidemiological and clinical research. Though randomized controlled trial is the gold standard for causal inference, most often we rely on observational studies due to ethical, logistical, and financial constraints [1]. Because of this, researchers frequently use observational data to evaluate treatment effects. Yet such studies are susceptible to bias due to unmeasured confounders [2], which affect both treatment assignment and outcomes and potentially distort the real effects [3].

Confounding is typically addressed by controlling for measured confounders through methods such as regression adjustment, propensity score matching or weighting, inverse probability weighting (IPW), standardization, or stratification [4, 5]. However, these methods cannot deal with unmeasured confounders, leaving uncertainty about the validity of the estimated causal effect. Here comes the idea of sensitivity analysis, which is the way to assess the robustness of the estimated effect to the unmeasured confounders [6].

VanderWeele and Ding [7] introduced E-value as a simple and accessible measure for sensitivity analysis in observational studies. It quantifies a minimum level of association that an unmeasured confounder should have with both treatment [8] and outcome to fully swipe away the observed treatment effect on outcome [7]. A large E-value suggests that a very strong unmeasured confounder should be there to nullify the observed effect, whereas a small value is susceptible to the robustness of that estimated effect to unmeasured confounders [9].

E-value is gaining popularity for its simplicity and minimum assumptions. Unlike many sensitivity techniques that require functional form, distribution, or prevalence of unmeasured confounders [10, 11], the E-value does not make such assumptions to work. This flexibility is the main reason to grow the E-value in different domains of research to explore the potential impact of unmeasured confounders, especially to the scientific and clinical audiences [8, 10, 11]. The original E-value methodology was developed for a single-time-point exposure-outcome setting [7]. We often deal with situations where treatments and confounders vary over time, while the outcome is measured at a single endpoint [5, 12]. For example, consider HIV treatment, where doctors modify the antiretroviral regimens multiple times based on how patients' CD4 counts and viral loads respond over months or years, and finally want to know whether the patient survives or develops AIDS by the end of follow-up [12]. Similarly, in cardiovascular research, clinicians adjust blood pressure or cholesterol medications repeatedly based on BP readings, while the outcome is whether someone has a heart attack or stroke at the end [13]. In a time-varying treatment confounder setting, only adjustment for baseline time-varying confounders is not sufficient and may introduce bias to the estimated effect [14]. Because time-varying confounders are affected by the prior treatment, this process is known as treatment-confounder feedback [5, 15]. Methods such as marginal structural models with IPW, parametric g-formula, and g-estimation have been developed to deal with this time-varying set-up in the presence of measured confounders [5]. But these advanced methods also rely on the assumption of no unmeasured confounding [16].

Therefore, there is a critical need for sensitivity analysis in longitudinal research with time-varying treatments and confounders [17]. E-value can be a flexible option that has not yet been extended broadly to this setting. By this, we could assess how much unmeasured confounding would be needed at any stage of follow-up to eliminate the observed treatment effect. This time-specific approach would reveal whether certain periods are more vulnerable to unmeasured confounding than others. Moreover, the conceptual simplicity that made the original E-value widely accessible [9] could be preserved in the longitudinal setting.

## Data Structure and Causal Framework

We consider a study with two time points where treatments and confounders vary over time, and the outcome is measured at the end of the follow-up time. Let $L_0$ is the baseline measured confounder, $A_0$ is the baseline treatment assigned after $L_0$ being measured, $L_1$ is the confounder being measured at time point 1, $A_1$ is the subsequent treatment, and $Y$ is the binary outcome measured at the end of the study. Additionally, let $U_0$ and $U_1$ represent unmeasured confounders at each time point that are not observed. The temporal ordering of these variables is: $L_0 \to A_0 \to L_1 \to A_1 \to Y$, where $L_1$ is affected by prior treatment and confounder: $A_0$ and $L_0$, respectively. Figure 1 represents the whole framework.

This two-time point structure serves as a foundation for understanding the more general case. The framework naturally extends to $T$ time points (treatment and confounders measured at $t = 0, 1, \ldots, T-1$, with outcome Y measured at time T) with sequential treatment-confounder feedback:

$$L_0 \to A_0 \to L_1 \to A_1 \to \cdots \to L_{T-1} \to A_{T-1} \to Y$$

where $\bar{L}_k = (L_0, \ldots, L_k)$ represents the measured confounder history through time $k$, $\bar{A}_k = (A_0, \ldots, A_k)$ represents the treatment history, and $\bar{U}_k = (U_0, \ldots, U_k)$ represents the unmeasured confounder history, and $k \leq T - 1$. While we present the two time point case for clarity and computational tractability, the methods extended here is applicable to the general $T$ time point setting.

We define $Y^{a_0, a_1}$ as the potential outcome that would be observed if treatments were set to $A_0 = a_0$ and $A_1 = a_1$. Our causal estimate is the risk ratio comparing the treatment regime "always treated" $(1,1)$ versus "never treated" $(0,0)$:

$$\text{RR} = \frac{P[Y^{1,1}]}{P[Y^{0,0}]}$$

## Identifying Assumption

We require the following assumptions to identify the causal effect from observational data.

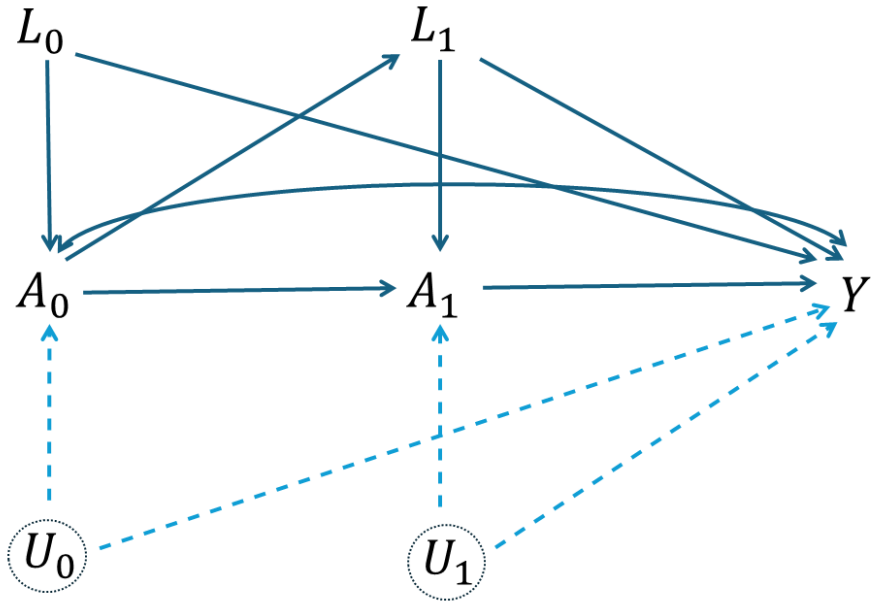

*Figure 1 Directed Acyclic Graph (DAG) illustrating the causal structure with time-varying treatments and confounders*

## Positivity

There will be a non-zero probability of receiving each treatment level for all measured confounders such that:

$$P(A_0 = a_0 \mid L_0 = l_0) > 0 \text{ for all } l_0 \text{ with } P(L_0 = l_0) > 0$$

$$P(A_1 = a_1 \mid L_0 = l_0, A_0 = a_0, L_1 = l_1) > 0 \text{ for all } (l_0, a_0, l_1)$$

It will ensure that we will observe individuals in all covariate-treatment combinations [18].

## Sequential Conditional Exchangeability

Treatment assignment is independent of the potential outcome given the treatment-confounder history at each time point t, with $t \in \{0,1\}$. That is

$$Y^{a_0, a_1} \perp A_0 \mid L_0$$
$$Y^{a_0, a_1} \perp A_1 \mid L_0, A_0, L_1$$

This sequential conditional exchangeability assumption would hold if there were no uncontrolled confounding [16].

## Consistency

The potential outcome under treatment level $a$ equals the observed outcome, when treatment level

equals to $a$:
$$A_t = a_t \Rightarrow Y^{a_0, a_1} = Y^{obs}$$
It allows us to equate the observed outcomes among those who received a certain exposure value to the potential outcomes that would be observed under the same exposure value [16, 19].

## Inverse Probability Weighting

We can estimate the causal effect using marginal structural models with IPW [20, 21]. The treatment models for two time points are defined as:
$$P(A_0 = 1 \mid L_0) = expit(\gamma_0 + \gamma_1 L_0)$$
$$P(A_1 = 1 \mid A_0, L_0, L_1) = expit(\beta_0 + \beta_1 A_0 + \beta_2 L_0 + \beta_3 L_1),$$
where $expit(x) = \frac{1}{1+e^{-x}}$ is the inverse logit function. To account for the time varying confounding affected by the prior treatment, we calculate the stabilized weight $SW_i$ as:
$$SW_i = \frac{P(A_0 = a_{0i})}{P(A_0 = a_{0i} \mid L_{0i})} \times \frac{P(A_1 = a_{1i} \mid A_{0i})}{P(A_1 = a_{1i} \mid A_{0i}, L_{0i}, L_{1i})}.$$
Finally, we fit a weighted logistic regression model to estimate marginal causal effects:
$$logit(P[Y = 1 \mid A_0, A_1]) = \alpha_0 + \alpha_1 A_0 + \alpha_2 A_1$$
weighted by $SW_i$. We then obtain the predicted marginal probabilities $\hat{P}(Y^{a_0, a_1})$ by setting $A^0 = a^0$ and $A^1 = a^1$ in the fitted model applying the inverse-logit transformation. The marginal risk ratio comparing always treated $(A^0 = 1, A^1 = 1)$ versus never treated $(A^0 = 0, A^1 = 0)$ is $\hat{P}(Y^{1,1})/\hat{P}(Y^{0,0})$, i.e., $\widehat{RR^{obs}} = \frac{\hat{P}(Y^{1,1})}{\hat{P}(Y^{0,0})}$.

## E-value Extension for Time Varying Confounding

The observed $\widehat{RR^{obs}}$ is valid until there is no unmeasured confounder. But the presence of $U_0$ and $U_1$ may cause bias and explain away the real effect. We aim to extend E-value framework to capture the required strength of unmeasured confounding to explain away the observed effect in presence of time varying treatment-confounders.

### Bias Factor for Single Time Point

Following VanderWeele and Ding [7], bias factor for unmeasured confounder at a single time point t is defined:
$$B_t = \frac{RR_{EU_t} \times RR_{UY_t}}{RR_{EU_t} + RR_{UY_t} - 1},$$
where $B_t$ denotes the amount by which an unmeasured confounder of the specific strength could alter the estimated effect, $RR_{EU_t}$ is strength of association between the exposure and the unmeasured confounder, and $RR_{UY_t}$ strength of association between the unmeasured confounder and the outcome at time $t$ [8]. The joint bounding factor, B, could take an infinite number of values depending on $RR_{EU_t}$, and $RR_{UY_t}$.

## E-value for Single Time Point

E value is the minimum strength of association that the unmeasured confounders should have with both the exposure and the outcome to explain away the estimated effect. For a particular time point t, it can be calculated as:

$$\text{E-value}_t = \widehat{RR_{\text{obs}}} + \sqrt{\widehat{RR_{\text{obs}}}(\widehat{RR_{\text{obs}}} - 1)},$$

where $\widehat{RR_{\text{obs}}}$ is the observed effect at that particular time. This $E$-value$_t$ can be derived from $B_t$ when $RR_{EU_t}$, and $RR_{UY_t}$ are equal for single time point setting [7, 8], which is one possible combination of $RR_{EU_t}$, and $RR_{UY_t}$.

## Combined Bias Factor for Multiple Time Points

For $T$ time points, the combined bias factor is:

$$B_{\text{total}} = B_0 \times B_1 \times \ldots \times B_{T-1}$$

$$= \frac{RR_{EU_0} \times RR_{UY_0}}{RR_{EU_0} + RR_{UY_0} - 1} \times \frac{RR_{EU_1} \times RR_{UY_1}}{RR_{EU_1} + RR_{UY_1} - 1} \times \ldots \times \frac{RR_{EU_{T-1}} \times RR_{UY_{T-1}}}{RR_{EU_{T-1}} + RR_{UY_{T-1}} - 1}$$

This multiplicative structure assumes that the biasing effect at each time point combined multiplicatively on the risk ratio scale, which is consistent with the risk ratio effect measures [22]. This assumption holds when the unmeasured confounding mechanisms at different time points do not interact in their effects on bias. This formula does not need $U_0$ and $U_1$ to be statistically independent, they may be correlated (e.g., if the same unmeasured factor persists over time, such as long-standing health behaviors, distance from home to hospital etc.) [23]. The multiplicative structure requires only that their contributions to bias combine multiplicatively, which is generally reasonable for confounding operating through separate time points [22].

The bias adjusted risk ratio is defined as:

$$RR^{\text{adj}} = \frac{\widehat{RR_{\text{obs}}}}{B_{\text{total}}} = \frac{\widehat{RR_{\text{obs}}}}{B_0 \times B_1 \times \ldots \times B_{T-1}}$$

This $RR^{\text{adj}}$ will be equal to 1 if

$$B_{\text{total}} = B_0 \times B_1 \times \ldots \times B_{T-1} = \widehat{RR_{\text{obs}}} \tag{1}$$

This $B_{\text{total}}$ value is actually denoting the least possible value of combined bias factor which can explain away the observed effect.

## E-values for Multiple Time Points

We consider three scenarios for reporting E-values at each time point following the framework of VanderWeele and Ding [7].

### Scenario 1: Equal Bias Distribution Across Time Points

When unmeasured confounding is equally strong at every time point $(B_0 = B_1 = ... = B_{T-1})$ then

$$B_t = \left(\widehat{RR_{\text{obs}}}\right)^{1/T} \tag{2}$$

And the E-value for each time point is:

$$E\text{-value}_t = \left(\widehat{RR_{\text{obs}}}\right)^{1/T} + \sqrt{\left(\widehat{RR_{\text{obs}}}\right)^{1/T} \left(\left(\widehat{RR_{\text{obs}}}\right)^{1/T} - 1\right)} \tag{3}$$

This represents the minimum strength when confounding is balanced equally across all time points.

### Scenario 2: All Confounding at a Single Time Point

Single time point confounding occurs when all bias factors except one equal 1. For instance, if $B_1 = ... = B_{T-1} = 1$, then $B_0 = \widehat{RR_{\text{obs}}}$, giving:

$$\text{E-value}_0 = \widehat{RR^{\text{obs}}} + \sqrt{\widehat{RR^{\text{obs}}}\left(\widehat{RR^{\text{obs}}} - 1\right)} \tag{4}$$

In a similar way, if $B_0 = B_2 ... = B_{T-1} = 1$, then E-value for time point 1 will be similar to the equation (4).

### Scenario 3: General Case

The first two cases are the special cases of this general one. In this approach, confounding may differ across times. For two timepoints, setting $B_1 = \frac{\widehat{RR_{\text{obs}}}}{B_0}$ allows visualization of confounding strengths at time point 1:

$$E\text{-value}_1 = \frac{\widehat{RR^{\text{obs}}}}{B_0} + \sqrt{\frac{\widehat{RR^{\text{obs}}}}{B_0}\left(\frac{\widehat{RR^{\text{obs}}}}{B_0} - 1\right)} \tag{5}$$

While this visualization becomes impractical for three or more time points, the constraint $B_{\text{total}} =$

$\widehat{RR}_{\text{obs}}$ defines the trade-off space mathematically for any T. To maintain both computational feasibility and practical interpretability, it is better to use equal bias distribution across time points, discussed in scenario 1. An important practical feature of this approach is its flexibility regarding estimation methods. In addition, the E-value calculations work identically whether the causal effect was estimated using g-computation [24], targeted maximum likelihood estimation [25], or any other valid approach [11].

In case of preventive exposure, observed risk ratio will be less than 1 [26]. For this situation, $\widehat{RR^{\text{obs}}}$ should be replaced with $\widehat{RR^{\text{obs}*}} = 1/\widehat{RR^{\text{obs}}}$ and run the whole analysis [7].

### Extension to Confidence Intervals

In addition to point estimate, we calculate E-value for the limit of the confidence interval (CI) closest to null value, 1 [7, 9]. E-value measure also can be used to assess the robustness of statistical inference to unmeasured confounding. If the CI does not include 1, then we calculate E-values using the confidence limit closest to the null. For a 95% CI $(L, U)$ with point estimate $\widehat{RR^{\text{obs}}} > 1$, if $L > 1$, we will use $L$ to calculate the E-value for each time point. If $\widehat{RR^{\text{obs}}} = 1$, it indicates that no confounding is needed to move the CI to include the null [7]. When $\widehat{RR^{\text{obs}}} < 1$, we will use the upper limit $U$ if $U < 1$.

The three scenarios for CIs will be mostly same with a little modification to those for the point estimate. From equation (1) to equation (5), we need to substitute $\widehat{RR^{\text{obs}}}$ by $L$ (in case of $\widehat{RR^{\text{obs}}} > 1$, and $L > 1$).

### E-values for Other Effect Measures for Ratio Scale

When outcome is relatively rare (<15%), the E-value formula mentioned earlier is directly applicable for Odds ratio (OR) or hazard ratio (HR). If the outcome is common, there need some approximations. For OR, $RR \approx sqrt(OR)$ should be used in the formal's equation (1) to equation (5). On the other hand, $RR \approx \frac{1 - 0.5^{\sqrt{HR}}}{1 - 0.5^{\sqrt{1/HR}}}$ should be used for HR. Detailed of these along with risk difference measure will be found in [7].

### Simulation Setup

I generated data for a group of 1000 individuals in the following manner:

- $U_0$ is sampled from a Bernoulli distribution with probability 0.4, representing an unmeasured baseline confounder.
- $L_0$ is sampled from a Bernoulli distribution with probability of 0.65, representing a measured baseline confounder.

- Conditional on $U_0$ and $L_0$, $A_0$ is drawn from a Bernoulli distribution with
$$P(A_0 = 1 \mid L_0, U_0) = expit(\alpha_0 + \alpha_1 L_0 + \alpha_2 U_0),$$
where $expit(x) = \frac{1}{1+e^{-x}}$ is the inverse logit function. Parameter values were taken as $\alpha_0 = -0.8, \alpha_1 = 1.2,$ and $\alpha_2 = 1$.
- $U_1$ is sampled from a Bernoulli distribution with probability 0.7, representing the unmeasured confounder at time point 1.
- Conditional on $A_0$ and $L_0$, $L_1$ is drawn from a Bernoulli distribution with
$$P(L_1 = 1 \mid A_0, L_0) = expit(\delta_0 + \delta_1 A_0 + \delta_2 L_0),$$
where $\delta_0 = -0.2, \delta_1 = 0.8,$ and $\delta_2 = 0.9$.
- Conditional on $A_0, L_0, L_1,$ and $U_1$, $A_1$ is drawn from a Bernoulli distribution with
$$P(A_1 = 1 \mid A_0, L_1, U_1) = expit(\xi_0 + \xi_1 A_0 + \xi_2 L_1 + \xi_3 U_1)$$
where $\xi_0 = -1.2, \xi_1 = 1,$ and $\xi_2 = 1.2,$ and $\xi_3 = 0.8$.
- Finally, given the values of $U_0, U_1, L_0, L_1, A_0,$ and $A_1$, potential outcomes $Y^{a_0, a_1}$ for all four treatment combinations $(a_0, a_1) \in \{(0,0), (0,1), (1,0), (1,1)\}$ are generated using a logistic model:
$$P(Y^{a_0, a_1} = 1) = expit(\mu_0 + \beta_0 a_0 + \beta_1 a_1 + \beta_2 L_0 + \beta_3 L_1 + \beta_4 L_0 L_1 + \gamma_0 U_0 + \gamma_1 U_1),$$
where $\mu_0 = -0.5, \beta_0 = 1.0, \beta_1 = 1.2, \beta_2 = 0.7, \beta_3 = 0.8, \beta_4 = 0.4, \gamma_0 = -0.7,$ and $\gamma_1 = -0.8$. The observed outcome Y is determined by consistency, $Y = Y^{A_0, A_1}$, corresponding to each individual's actual treatment history. The specific R code to reproduce this simulation and the results has been provided in supplementary file.

## Results

Following he simulation mechanism, we found the causal risk ratio by comparing always treated vs never treated was 1.87. After excluding the unmeasured confounders from baseline and 1st time point, we found the observed causal risk ratio $\widehat{RR}^{obs} = 1.73$. The corresponding 95% non-parametric bootstrap confidence interval (CI) was found to be (1.52, 1.99).

### E-values for Point Estimate

We calculated the E-value to assess the probable strength of the unmeasured confounder needed to explain away the observed effect. For scenario 1, we got the E-value for each time point as 1.96. This value indicates that an unmeasured confounder associated with both treatment and outcome by a risk ratio of 1.96-fold at each time point would be required to explain away. If unmeasured confounding presents at a single time point (scenario 2) instead of twos, 2.85-fold associations at any time point (baseline or follow-up) to fully explain away the effect.

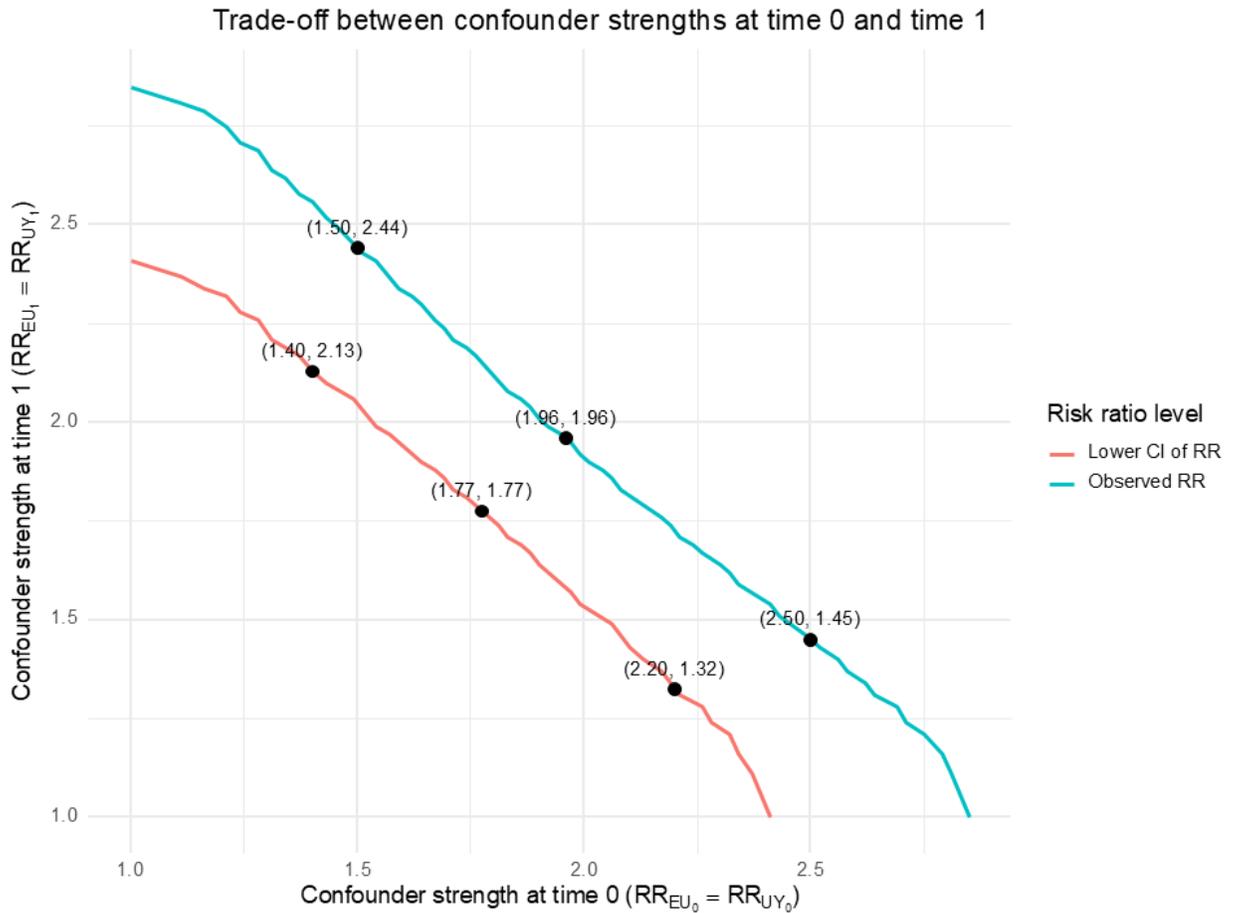

*Figure 2 Strength of unmeasured confounder with treatment and outcome at each time point*

For general case, we could have a combination of $(RR_{EU_0}, RR_{UY_0})$, $(RR_{EU_1}, RR_{UY_1})$ pairs which could swipe away the observed effect completely. Figure 2 represents such all-possible combinations. For example, if unmeasured confounding at baseline has association of 1.50-fold with both treatment and outcome, then confounding at follow-up would need 2.44-fold associations with treatment and outcome to completely nullify the observed risk ratio.

### E-values for Confidence Interval

Beyond assessing the point estimate's robustness, we examined how vulnerable our statistical inference is to unmeasured confounding. Using the lower confidence limit $L = 1.52$, we assessed how much unmeasured confounding it would take not to eliminate the effect entirely, but simply to shift our confidence interval to include the null value to make the result non-significant. For scenario 1, we got the equal E-value (1.77) for each time point. It indicates, unmeasured confounding with associations of just 1.77-fold at each time point could push our confidence interval to include RR = 1.0.

If unmeasured confounding occurs through a single time point (either at baseline or 1st time point) 2.41-fold associations would suffice to make our finding non-significant. For the general case with confidence intervals, we have multiple combinations of ($E$-value$_{0,CI}$, $E$-value$_{1,CI}$) pairs that could shift the confidence interval to include the null. Figure 2 represents all such possible combinations. For example, if unmeasured confounding at baseline has associations of 1.40-fold with both treatment and outcome, then confounding at follow-up would need 2.13-fold associations with treatment and outcome to make the result statistically non-significant.

## Application to Real Data: Insulin Resistance and Cardiovascular Disease

To demonstrate how this works in practice, we applied our E-value extension to a recently published paper by Feng et al. [27], which examined insulin resistance trajectories and cardiovascular disease risk using data from a large Chinese cohort in OR scale. Since the outcome is less than 15%, E-value formula can be directly applicable here [7].

### Study Overview

CHARLS is a nationally representative longitudinal cohort of Chinese adults aged 45+ years, publicly available at https://charls.pku.edu.cn/. The analysis included 3,966 participants with blood samples collected at baseline (2011, Wave 1) and follow-up (2015, Wave 3), with cardiovascular disease (CVD) outcomes assessed through 2018 [27]. This paper has several outcomes for different models. I will focus mainly on the result of model 5.

They defined the time varying exposure as Triglyceride-glucose body mass index (TyG-BMI), dichotomized at median, measured at Time 0 (2011) and Time 1 (2015). Age, sex, residence, and education were considered as time invariant confounders. Smoking, alcohol consumption, marital status, chronic conditions (hypertension, dyslipidemia, diabetes, kidney and liver disease), laboratory parameters (HDL, LDL, high-sensitivity C-reactive protein) were considered as time dependent confounders in this study. Lastly, the outcome was the incident of CVD (heart disease or stroke: binary variable) assessed in 2018.

Participants were classified into four trajectory groups: low stable (0,0), increasing (0,1), decreasing (1,0), and high stable (1,1). Using longitudinal targeted maximum likelihood estimation (LTMLE), the study found that sustained elevated insulin resistance (high stable trajectory) was associated with increased CVD risk compared to low stable levels. They reported an OR of 1.38 (95% CI: 1.07 - 1.77, p = 0.013). The authors also calculated E-value of 2.093 and concluded the association was "moderately robust to unmeasured confounding" [27].

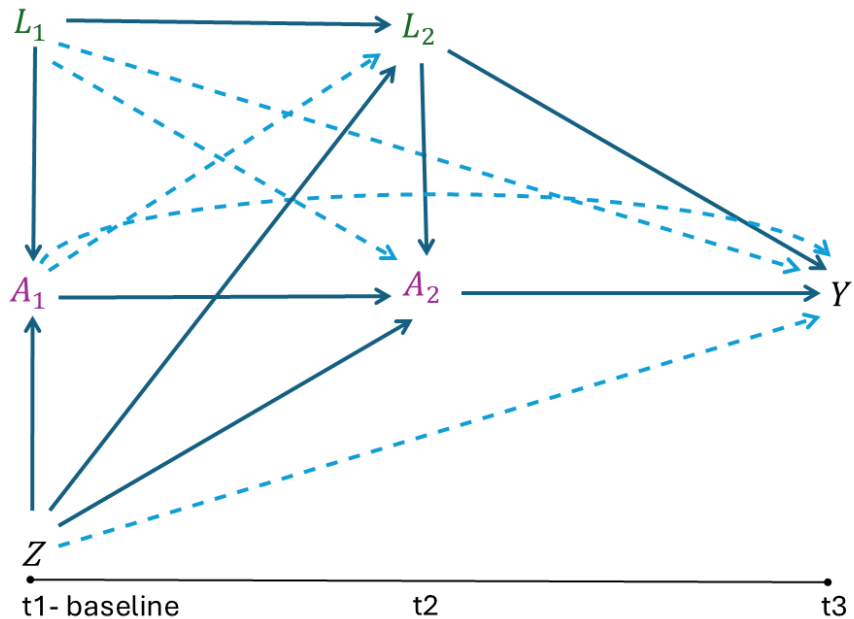

Figure 3 Assumed Directed Acyclic Graph (DAG) illustrating sequential relationships among exposure, outcome, and time-varying confounders across multiple time points. Z refers to the baseline covariates; L refers to the time-varying covariates; A refers to the exposure, including TyG-BMI and TyG index; 3 time points were included, including t1 (in 2011), t2 (in 2015), and t3 (in 2018) [27]

For the sensitivity analysis, they used the E-value just like for a single time point. But there should be more possibilities. For example, if the confounding works equally at the two-time points, then

$$B_0 = B_1 = \sqrt{1.375} = 1.173 \approx 1.17$$

So, the E-value for each time point will be 1.63 (from equation (3)). It means, an unmeasured confounder would need 1.63-fold associations with both insulin resistance and CVD at each time point to explain away the effect. This is notably lower than the single-timepoint E-value, suggesting more vulnerability than initially apparent. However, there is a possibility of different combinations of the strength of the unmeasured confounding in two points. Figure 4 presents such combinations.

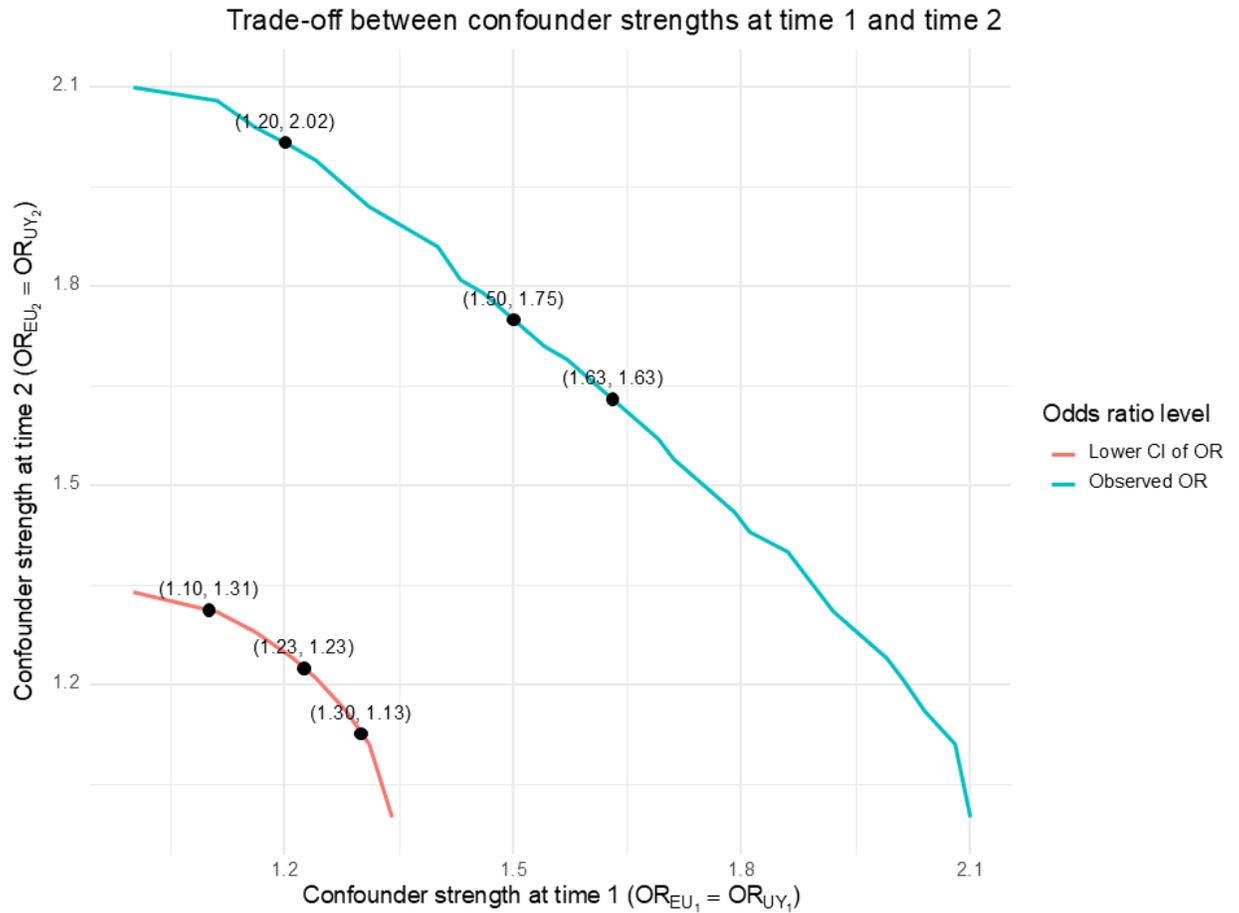

*Figure 4 Combinations of unmeasured confounding strengths with both treatment and outcome across time points required to nullify the observed effect*

For example, if unmeasured confounding of 2.02 fold association present in 1st time point, only confounding of 1.20 fold association would need to explain away the observed effect. In the similar way, unmeasured confounding with associations around 1.23, and 1.34 fold could push the confidence interval to include the null, making the result non-significant if equal confounding at two time points and confounding at only one time point is being considered, respectively. There is also a possible combination of strength of unmeasured confounding to make the confidence interval to include the 1. For instance, confounding of 1.30-fold association at 1st time point and 1.13-fold association at 2nd time point could push the confidence interval towards the null value.

### Discussion

This study extends the E-value framework to longitudinal settings with time-varying treatments and confounders. The extension reveals that unmeasured confounding operating across multiple time points requires less strength at each individual time point to explain away an observed effect. In the simulation presented, where a single-timepoint approach would suggest an E-value of 2.85, the time-varying framework indicates that unmeasured confounders with 1.96-fold associations at each time point could nullify the observed effect when confounding is equally distributed. This suggests that many published longitudinal studies may be more vulnerable to unmeasured confounding than their reported single-timepoint E-values indicate [27].

This extended approach builds on the foundation by quantifying the specific combinations of unmeasured confounder strengths that would be required to explain away observed effects. Importantly, the multiplicative structure does not require statistical independence between unmeasured confounders at different time points, accommodating realistic scenarios where the same underlying factor persists across the study period [23]. There are several recently proposed method for sensitivity analysis in longitudinal set-up, but they often require specifying parametric models for unmeasured confounders or making assumptions about their distributions [28, 29]. While these approaches can provide more nuanced assessments under certain conditions, they can be challenging to implement and difficult to communicate to the applied researchers and clinical audiences. The E-value framework, in contrast, requires no assumptions about the functional form or distribution of unmeasured confounders, making it more accessible for widespread use [7, 9]. This extension preserves this accessibility while accommodating the complexity of time-varying treatment-confounder settings.

The three reporting scenarios: equal bias distribution, single-timepoint confounding, and general case, provide flexibility in interpretation. The equal-bias distribution scenario is recommended as the primary reporting strategy because it offers a single, interpretable metric for each time point. Researchers should interpret this as a lower bound: studies are at least this vulnerable to unmeasured confounding, potentially more so if confounding operates at unmeasured intermediate occasions. The single-timepoint E-value can be reported secondarily for comparison with existing literature [6, 7], while the general case visualization allows exploration of specific confounding scenarios when the number of timepoints is small (less than or equal to 3).

The two-timepoint framework demonstrates the core methodological contribution while maintaining computational and interpretive clarity. The mathematical extension to $T$ time points is straightforward, though visualization and practical implementation for K > 3 require additional development. Our approach establishes the theoretical foundation and provides immediately applicable tools for common longitudinal study designs [12, 20].

## Conclusion

This work extends the E-value framework to longitudinal settings with time-varying treatments and confounders, providing researchers with accessible tools to assess the robustness of causal estimates to unmeasured confounding. The proposed approach demonstrates that vulnerability to unmeasured confounding in time-varying settings can differ substantially from single-timepoint assessments, with important implications for interpreting findings from longitudinal studies. By offering three reporting scenarios: equal bias distribution, single-timepoint confounding, and the general case, this extension maintains the simplicity and minimal assumptions that make E-values widely applicable while accommodating the complexity of time-varying treatment-confounder relationships. As longitudinal observational studies become increasingly prevalent in epidemiological research, these methods provide a practical framework for transparent sensitivity analysis that can strengthen causal inference and guide the interpretation of research findings.

## Strength & Limitation

This work addresses a notable gap by extending the E-value framework, originally designed for single time-point settings, to complex longitudinal setups with time-varying treatments and confounders with several scenarios. Another mentionable point is that, this approach does not rely on any distribution or parametric form of unmeasured confounders.

In contrast, we have some limitations. Firstly, the combined bias factor assumes that biasing effects at each time point combine multiplicatively. While reasonable for many risk ratio measures, it may not capture complex non-multiplicative interactions between different confounding mechanisms. Secondly, researchers applying this framework to studies with many time points (> 3) should interpret the equal-distribution E-value as a lower bound on vulnerability.

## Future Work

Several extensions of this work warrant investigation. First, while we focused on binary exposures and outcomes, extensions to continuous exposures or time-to-event outcomes would broaden applicability. Second, the interaction between time-varying confounding and effect modification deserves attention, as unmeasured effect modifiers could create patterns of confounding that vary across subgroups. Finally, methods for incorporating prior knowledge about the likely magnitude of unmeasured confounding from external sources or validation studies could strengthen sensitivity analyses.

**Abbreviations**: DAG, Directed Acyclic Graph; RR, Risk Ratio; CI, Confidence Interval; SUTVA, Stable Unit Treatment Value Assumption; MSM, Marginal Structural Models; IPW, Inverse Probability Weighting.